\documentclass[aps,prb,twocolumn,showpacs]{revtex4}
\usepackage{graphicx}
\usepackage{amssymb}

\begin{document}


\title{Magnetic and thermodynamic properties of $R$MnO$_3$ ($R$ = Pr, Nd)}

\author{J. Hemberger$^1$, M. Brando$^1$, R. Wehn$^1$, V.Yu.~Ivanov$^2$, A.A.~Mukhin$^2$,
A.M.~Balbashov$^3$, and A. Loidl$^1$}
\affiliation{%
$^1$Experimentalphysik V, Elektronische Korrelationen und
Magnetismus, \\
Institut f\"{u}r Physik, Universit\"{a}t Augsburg, D-86135 Augsburg, Germany \\ %
$^2$ General Physics Institute of the Russian Academy of Sciences, 119991 Moscow, Russia \\ %
$^3$ Moscow Power Engineering Institute, 105835 Moscow, Russia  %
}

\begin{abstract}
The ground state properties of the pure perovskite compounds
PrMnO$_3$ and NdMnO$_3$ were investigated by magnetization,
magnetic AC susceptibility and specific heat measurements. A
strongly anisotropic behavior has been detected for temperatures
below the antiferromagnetic phase transition $T_{\rm N}\approx
100$~K.
 The susceptibility and the weak spontaneous
ferromagnetic moment appear to be different in both compounds due
to different anisotropic rare earth contributions.
The specific heat shows strong Schottky type contributions at low
temperatures, which for NdMnO$_3$ strongly depend on the magnetic
field. A spin reorientation phase transition (spin-flop type)
induced by a magnetic field along $b$ axis was observed in
NdMnO$_3$ at $H\sim 110$~kOe and $T=5$~K. All results can
consistently be explained by anisotropic contributions of the rare
earth ions: In PrMnO$_3$ the electronic ground state is determined
by a low lying quasidoublet split by the crystal field ($\sim
19$~K). In NdMnO$_3$ the Kramers doublet is split by an exchange
Nd-Mn field ($\sim 20$~K).
\end{abstract}


\maketitle


In recent years perovskite-type manganites have attracted
considerable interest particularly due to the appearance of
complex magnetic and transport properties, such as the colossal
magneto resistance effect (CMR).\cite{1} 
In these materials the physical properties are determined by a
delicate interplay of charge, spin, orbital, and lattice degrees
of freedom.\cite{3} Substituting the A-site trivalent rare earth
($R$) by divalent Sr or Ca  ($R_{1-x}A_x$MnO$_3$) changes the
filling of the 3$d$ Mn bands, introducing double-exchange
interactions and thereby reducing Jahn-Teller type distortion and
orbital order.\cite{par00b} In addition doping also changes the
tolerance factor, which reflects deviations from the ideal
perovskite structure.\cite{ima98} The Mn-O-Mn bond angle can be
altered and the super-exchange between the Mn sites is influenced.
As a result the perovskite manganites exhibit a rich phase diagram
including different types of ordering phenomena and ground states
like charge order, orbital order 
and antiferromagnetic (AFM) or ferromagnetic (FM) insulators or
metals.\cite{ima98}

Focusing on the parent compounds, the undoped manganites, the
question arises, how, in addition to a small change in the
tolerance factor, the specific choice of the $R$ ion may alter the
magnetic properties. Pure LaMnO$_3$ exhibits a Jahn-Teller
distorted orthorhombic structure below $T \approx 800$~K and
changes from an insulating paramagnet to an insulating
antiferromagnet at $T_{\rm N} \approx 140$~K .\cite{wol55,jon50}
Substituting the nonmagnetic trivalent La by isovalent Pr or Nd,
the electronic configuration of the Mn ions is qualitatively
conserved and an orbitally ordered 
A-type AFM ground state is established.\cite{wu00,mun00,quez68}
But how do the magnetic degrees of freedom of the Pr$^{3+}$ and
Nd$^{3+}$ interact with the ordered Mn$^{3+}$ sublattice and
contribute to magnetic properties of the manganites~?  The purpose
of this work is to shed some light on the role of the magnetic $R$
ions and the mechanisms which determine the low temperature
magnetic and thermodynamic properties in the pure perovskite
manganites.


The single crystalline samples investigated in this work have been
prepared by zone melting as described elsewhere.\cite{balb} In
contrast to LaMnO$_3$ which is rather sensitive to growth
conditions and thermal processing parameters, PrMnO$_3$ and
NdMnO$_3$ reveal a much weaker dependence of their properties on
these parameters indicating a higher stoichiometric stability.
X-ray diffraction pattern, taken at room temperature, revealed a
single-phase orthorhombic Pbnm structure with the lattice
constants $a=5.448(6)$, $b=5.822(6)$, $c=7.582(0)$ for PrMnO$_3$
and $a=5.415(9)$, $b=5.837(9)$, $c=7.548(1)$ for NdMnO$_3$.
These data are in very good agreement with the data of
stoichiometric samples described in literature.\cite{Cherepanov} A
comparison of our results to studies of the dependence of the
lattice parameters in compounds with non-isovalent impurities or
oxygen deficiency\cite{Kitayama, Kamehashira, Jirak}
gives an upper limit of the defect of our samples of 0.5~\%.
 Another evidence for the quality of our
samples is the Neel temperature which for NdMnO$_3$ and PrMnO$_3$
refers to the highest cited in literature. It is known that
$T_{\rm N}$ is significantly reduced upon the smallest deviations
of the nominal stoichiometry.\cite{Ritter, Jirak}

Laue diffraction was utilized to orient the samples with respect
to the pseudo-cubic crystallographic axis. The magnetization
measurements were performed using a commercial SQUID system (MPMS,
{\sc QuantumDesign}) between 1.8~K and 400~K and in magnetic
fields up to 50~kOe. Additional measurements were performed
employing an extraction magnetometer and an AC susceptometer ({\sc
Oxford} Teslatron) in fields up to 140~kOe. The specific heat has
been measured on crystals with masses $m \lesssim 20$~mg at
temperatures 5~K~$\leq T \leq$~250~K with a noncommercial setup
utilizing an AC method.

\begin{figure}[tb]
\includegraphics[clip,width=65mm]{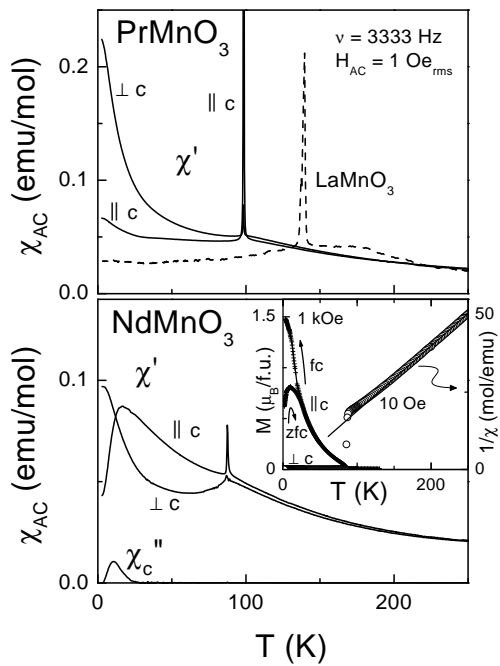}
\caption{Real part of the AC susceptibility $\chi_{\rm AC}'$ of
PrMnO$_3$ (upper frame) and NdMnO$_3$ (lower frame) measured along
and perpendicular to the crystallographic $c$ direction. The
dashed line in the upper frame represents $\chi_{\rm AC}$ of
LaMnO$_3$. For NdMnO$_3$ the imaginary part $\chi_{\rm AC}''$
(loss) along c is shown as well. Inset: Reciprocal DC
susceptibility (open circles, right scale) together with
corresponding Curie-Weiss fit (straight line) and
field-cooled/zero-field-cooled magnetization (crosses, left scale)
of NdMnO$_3$. \label{susceptibility}}
\end{figure}

The temperature dependent measurements of the magnetic AC
susceptibility are shown in Fig.~\ref{susceptibility}. The AFM
transitions show up as sharp peaks at $T_{\rm N}=99$~K for
PrMnO$_3$ and $T_{\rm N}=88$~K for NdMnO$_3$, respectively. For
comparison the data as obtained in a twinned LaMnO$_3$ single
crystal\cite{par00} with $T_{\rm N}=139$~K are also shown (dashed
line in the upper frame of Fig.~\ref{susceptibility}).
In the magnetically ordered regime of the $R$ compounds a
significant anisotropy shows up. The values of $\chi'$ in
PrMnO$_3$ and NdMnO$_3$ are much higher than in LaMnO$_3$, at
least in the $a,b$ plane. This obviously results from additional
magnetic contributions of the $R$ ions. At higher temperatures in
the paramagnetic regime these contributions can easily be
considered by adding the respective effective paramagnetic moment
$\mu_{\rm eff}$. A Curie-Weiss fit to the susceptibility of
NdMnO$_3$, shown in the inset of Fig.~\ref{susceptibility}, yields
$\mu_{\rm eff} = 6.16$~$\mu_{\rm B}$ and $\Theta_{\rm CW}\approx
10$~K. Experimental values for LaMnO$_3$ are $\mu_{\rm eff} =
5.10$~$\mu_{\rm B}$ and $\Theta_{\rm CW}\approx 66$~K.\cite{par00}
The expected effective moment of $\mu_{\rm eff} = 6.09$~$\mu_{\rm
B}$ for the superposition of the Nd (4$f^3$) and the Mn (3$d^4$)
sublattice agrees well with the experimental findings. Similar
results were obtained for PrMnO$_3$.

To understand the enhanced anisotropic contributions below $T_{\rm
N}$ one has to focus on the electronic ground state of the $R$
ions, which via spin-orbit coupling depends on the local symmetry
of the crystal field (CF) and on the exchange interaction with the
Mn sublattice.
The Pr compound reveals a strong increase of the susceptibility
parallel to the $a,b$ plane towards low temperatures. Along the
$c$ direction $\chi$ remains almost constant, comparable to the
behavior in LaMnO$_3$. The AC susceptibility resembles the results
reported from DC measurements.\cite{mukh01} In NdMnO$_3$ the
susceptibility of the $a,b$ plane is slightly smaller than in
PrMnO$_3$, but below $T_{\rm N}$ and along $c$, $\chi'$ rises with
decreasing temperature, reveals a cusp at $T\approx 18$~K and
decreases towards lower temperatures. The drop off in the real
part of the susceptibility of NdMnO$_3$ at about 10~K is
accompanied by a peak in the imaginary part $\chi"$. In addition a
distinct splitting between the DC magnetization measured under
field-cooled (FC) and zero-field-cooled (ZFC) conditions can be
detected below a splitting temperature $T_{\rm FC/ZFC}$ as shown
in the inset of Fig.~\ref{susceptibility}. On increasing magnetic
dc field the splitting is shifted from temperatures close to
$T_{\rm N}$ down to $T_{\rm FC/ZFC}\approx 22$~K for $H=1$~kOe.
This type of nonergodic behavior seems to be connected with the
freezing of the domain dynamics within the ordered Mn sublattice.
Nevertheless, the high values of the FC magnetization of up to
1.5~$\mu_{\rm B}/$f.u. cannot be explained by the contribution of
the Mn spins alone, because these are ordered in an A-type AFM
structure, as discussed below. These phenomena, namely the FC/ZFC
splitting and the loss features of the AC susceptibility, which
appear only along the $c$ direction, have rather to be interpreted
in terms of domain freezing processes and the corresponding
longitudinal relaxation of the Nd sublattice, polarized by the
Nd-Mn exchange coupling.

\begin{figure}[tb]
\includegraphics[clip,width=65mm]{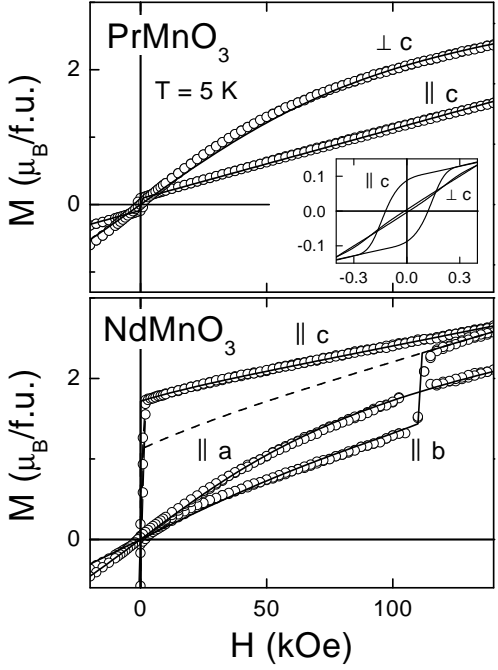} \caption{Hysteresis loops of
PrMnO$_3$ (upper frame) and NdMnO$_3$ (lower frame) measured for
different orientations with respect to the magnetic field at
$T=5$~K. Lines denote fits employing the model described in the
text. \label{hysterese}}
\end{figure}

High values of the FM component along c can also be detected in
the hysteresis loops at $T=5$~K, shown in Fig.~\ref{hysterese}.
The lower frame of Fig.~\ref{hysterese} shows the magnetization
for NdMnO$_3$ up to 140~kOe. Along the $c$ axis a spontaneous FM
moment of $M_s\approx1.7$~$\mu_{\rm B}$ is detected. The
corresponding value for PrMnO$_3$, $M_s\approx0.09$~$\mu_{\rm B}$
(see inset of Fig.~\ref{hysterese}) is by a factor of $\sim$20
smaller and comparable to the findings for LaMnO$_3$.\cite{par00}
 In PrMnO$_3$ and LaMnO$_3$ the
small FM component arises from the the canting of the A-type AFM
structure due to Dzyaloshinsky-Moriya interaction, which is
generated by a buckling of the orthorombic oxygen
octahedra.\cite{pim00}
  This explanation does not hold for the high
values of $M_s$ found in NdMnO$_3$. A FM ground state of the Mn
sublattice can be ruled out. Hole doping $R_{1-x}A^{2+}_x$MnO$_3$
leads to a FM ground state only for $x\gtrsim 0.1$.\cite{ima98}
Also the jump in the magnetization at $H\approx 110$~kOe denotes a
field induced spin-reorientation transition (spin-flop
type).\cite{note1} This effect can only be explained by the
reorientation of the spins within an AFM spin structure. Thus the
FM component along the $c$ axis in NdMnO$_3$ has to be ascribed to
the Nd spins. The Nd sublattice is polarized due to an effective
exchange field of the Mn sublattice, which exhibits a small FM
moment in $c$ direction. At low temperatures the Nd spins are
almost completely aligned in the internal exchange field, which is
indicated by the drop of $\chi_c$. In contrast the effective field
due to the FM component of the Mn lattice in $c$ direction does
not couple to the Pr spins. In PrMnO$_3$ an enhanced
susceptibility as well as a spontaneous magnetization along $c$
are absent.


To describe the observed experimental findings and to determine
the ground state characteristics of the $R$ ions one can utilize
the models already used in the $R$ orthoferrites having the same
crystal structure, however G-type AFM ordering.\cite{whit69} In
$R$MnO$_3$ the ground multiplet of the $R$ ions occupying low
symmetry sites (point group $C_s$) is split by a CF into singlets
for non-Kramers ions (Pr$^{3+}$) and into doublets for Kramers
ions (Nd$^{3+}$). One arrives at an energy level scheme where the
low-lying state is split due to crystal, external, and  exchange
fields. The exchange fields are determined by the isotropic and
anisotropic $R$-Mn exchange, proportional to the FM and AFM
vectors $\bf F=(M_{1}+M_{2})$$/2M_{0}$ and $\bf
A=(M_{1}-M_{2})$$/2M_{0}$, respectively. There $\bf M$$_{1,2}$
denote the Mn sublattice magnetization of adjacent ferromagnetic
layers with saturation value $M_{0}$. For the Pr system the
splitting of the quasidoublet
(i.e. the two lowest CF split singlets) is of the form %
\begin{equation}
\label{eqn-Pr} \Delta^\pm \approx 2\left[ \Delta _{\rm cf}^{2}+
\left( {\Delta _{z}+ {\bf \mathbf{\mu}^\pm H}}\right)^{2}
\right]^{1/2} ,
\end{equation}
where 2$\Delta _{\rm cf}$ and 2$\Delta _{z} = 2\Delta _{z}^0A_{z}$
are the splittings by the CF and Pr-Mn exchange
interactions\cite{note2}, and ${\mathbf{\mu}^\pm} =
\left(\mu_x,\pm\mu_y,0 \right)$ is the magnetic moment of the
ground state which lies in the $a,b$ plane. $x,y,z$ correspond to
$a,b,c$ axis, respectively, and the signs $\pm$ correspond to the
two non-equivalent $R$ positions. Here we have taken the observed
magnetic anisotropy of the Pr contribution into account, namely
the absence of a FM moment along the $c$ axis, implying that the
moments of the Pr ions are oriented within the $a,b$ plane.

Taking into account the local symmetry of the g-factor and the
exchange field at the A-sites \cite{whit69} the splitting of the
Nd$^{3+}$ Kramers doublet can be expressed through $\bf A$ and
$\bf H$ by
\begin{eqnarray}
\nonumber \Delta^\pm & \approx & 2 [ \left( \mu _{z}{H_{z}}
 + \Delta_{y} \pm \Delta_{x} \right)^{2} + \Delta_{z}^2
 + 2\mu_y'H_y \Delta_{z} + \mu_y^2 H_y^2 \\
\label{eqn-Nd} & &  + \mu_x^2 H_x^2 \pm 2\mu_{xy}^2H_xH_y \pm
2\mu_x'H_x \Delta_{z}
 ]^{1/2},
\end{eqnarray}
where $2\Delta_{\alpha} = 2\Delta _{\alpha}^0 A_{\alpha}$ (with
$\alpha =x, y, z$) are the Nd-Mn exchange splittings and
$\mu_{x,y,z,xy}$ and  $\mu_{x,y}'$ are determined
by the g-factor of the doublet and 
the Nd-Mn exchange parameters, respectively. %
Thus, in NdMnO$_3$ the splitting of the ground state is basically
driven by the Nd-Mn exchange interaction, while in PrMnO$_3$
exchange splitting is absent in the $F_zA_y$ configuration, but
instead the CF splitting dominates.

The magnetic properties
were analyzed using the non-equilibrium thermodynamic potential
\begin{eqnarray*}
\Phi\left({\bf H,A}\right) &=& \Phi_{\rm Mn}({\bf H,A}) \\
 & & -\frac{1}{2}Nk_{\rm B}T
 \sum_{i=\pm}\ln\left(2\cosh \left(\frac{\Delta_i}{2k_{B}T}\right)\right),
\end{eqnarray*}
where the corresponding potential of the Mn subsystem $\Phi_{\rm
Mn}(\bf H,A)$ was considered phenomenologically.\cite{pim00} The
magnetization ${\bf M} = -
\partial \Phi / \partial \textbf{H}$ is defined by
\[
M_x = \frac{2N \mu_x^2H_x}{\Delta}
 \tanh\left(\frac{\Delta}{2k_{B}T}\right)
 + \left(\chi_{\perp} + \chi _{x}^{\rm VV}\right)H_{x} 
\]
for $\textbf{H} \parallel a$ and by
\[
M_z = m_z^{\rm Mn} + N \mu_z
 \tanh\left(\frac{\Delta}{2k_{\rm B}T}\right)
 +(\chi_{\parallel} + \chi_z^{\rm VV})H_z
\]
for $\textbf{H} \parallel c$ with $\mu_{z}$=0 for PrMnO$_3$. Here
$m_{z}^{\rm Mn}$ is the weak ferromagnetic moment, $\chi _{\bot}$
the perpendicular susceptibility of the Mn subsystem, and $\chi
_{x,z}^{\rm VV}$ is the Van-Vleck susceptibility of the $R$ ions
due to excited states. In the case of the spin-flop transition
along the $b$ axis in NdMnO$_3$ the low and the high field regime
have to be treated separately. The results of the simulation for
the magnetization curves are shown as solid lines in
Fig.~\ref{hysterese} and the corresponding main parameters of the
$R$ ions are given in Tab.~\ref{schottky} and are in a good
agreement with similar data evaluated recently from temperature
dependent magnetization measurements \cite{mukh01} and
submillimeter optical studies.\cite{mukh03} A more detailed
analysis of the spin reorientation and high-frequency magnetic
excitations of NdMnO$_3$ will be presented in a separate
publication.


\begin{figure}[tb]
\includegraphics[clip,width=65mm]{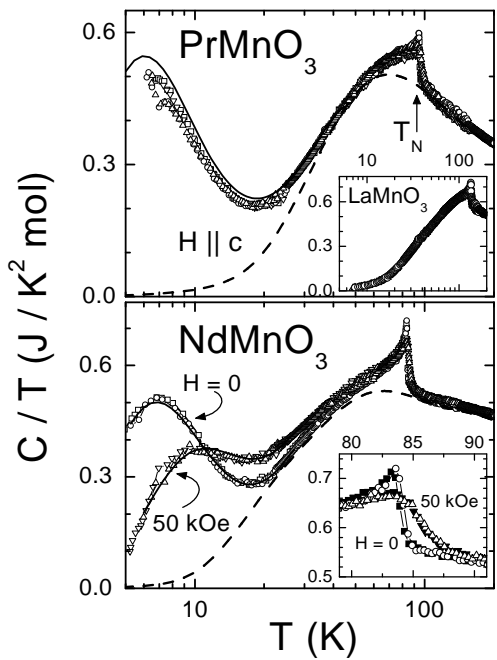}
\caption{Temperature dependence of the specific heat of PrMnO$_3$
(upper frame) and NdMnO$_3$ (lower frame) presented as $C/T$
versus temperature on a logarithmic scale in external magnetic
fields of 0~kOe (squares) and $H=50$~kOe (triangles). The solid
lines are calculated employing a two-level Schottky term as
described in the text. Dashed lines display the lattice
contribution. The inset in the upper frame shows the data for
LaMnO$_3$ (non magnetic $R$). The inset of the lower frame
displays the specific heat of NdMnO$_3$ near $T_{\rm N}$.
\label{CT}}
\end{figure}

To get an independent measure of the above determined values of
the splitting $\Delta$, we performed measurements of the specific
heat in PrMnO$_3$ and NdMnO$_3$ in the temperature range from 5 to
250~K and in magnetic fields up to 50~kOe (see Fig.~\ref{CT}).
The AFM transitions of both compounds show up as lambda-like
anomalies which are smeared out in an external field (see the
inset of the lower frame). A temperature hysteresis (open vs
closed symbols) cannot be observed, neither at zero field nor at
50~kOe, indicative of a second-order transition. At low
temperatures (below 20~K) in $C/T$ a broad maximum appears in
both, PrMnO$_3$ (upper frame of Fig.~) and NdMnO$_3$ (lower
frame). For comparison also the specific heat of LaMnO$_3$ was
examined (see inset of the upper frame). Here this electronic
contribution at low temperatures is absent.

Even if it is difficult to calculate and separate the specific
heat of the lattice at elevated temperatures, we tried to
parameterize the phonon contribution utilizing a Debye term
($\Theta_D\approx285$~K) and two Einstein terms related to the
minimal and maximal phonon energies in the
perovskites.\cite{may00} The results are shown as dashed lines in
Fig.~. In addition a Schottky contribution for a two-level system
(doublet) of the form $C \sim \Delta^2/(T\cosh(\Delta /2k_{\rm
B}T))^2$ was
used to fit the low temperature anomaly (solid lines in
Fig.~\ref{CT}). The corresponding values for the splitting
$\Delta$ are given in Tab.~\ref{schottky}. In zero external field
the results are in excellent agreement with those obtained from
the magnetization data. As given by Eq.~(\ref{eqn-Pr}) in the Pr
system the splitting $\Delta$ is not changed by an magnetic field,
while it is increased for NdMnO$_3$. According to
Eq.~(\ref{eqn-Nd}) one can estimate $\mu_c$ of the Nd ions from
the difference between $\Delta(H_c$=0) and $\Delta(H$=50kOe),
which again meets the first result (see Tab.~\ref{schottky}).
These results are in accordance with optical measurements in the
submillimeter regime.\cite{mukh03}

\begin{table}[b]
\caption{Parameters gained from the fitting of the magnetization
data (Fig.~\ref{hysterese}) and the Schottky contributions to the
specific heat (Fig.~\ref{CT}) \label{schottky}}
\begin{tabular}
{l|cc}%
\hline \hline
& ~~~~~~~~Pr~~~~~~~~ & ~~~~Nd~~~~ \\ \hline %
$\Delta(H_c=0$) [K] from $C_p$ & 19.3 $\pm$ 1 & 21.6 $\pm$ 1 \\
$\Delta(H_c=0$) [K] from $M$& $18.7\pm 1$ &
$20.2\pm 1$ \\
$\Delta(H_c=50$~kOe) [K] from $C_p$ & 19.3 $\pm$ 1 & 32.5 $\pm$ 1 \\
$\mu_c$ [$\mu_{\rm B}$] from $C_p$ &  &  1.6 $ \pm  0.1$\\
$\mu_{c}$ [$\mu_{\rm B}$] from $M$ &  & 1.9 $\pm 0.1$ \\
$\mu_{\perp c}$ [$\mu_{\rm B}$] from $M$ & 2.1 $\pm 0.1$ & ($a$) 1.8 $\pm 0.1$ \\
& & ($b$) 1.2 $\pm 0.1$\\
\hline \hline
\end{tabular}
\end{table}


In conclusion, we presented measurements of the magnetic
susceptibility, magnetization and the specific heat of NdMnO$_3$
and PrMnO$_3$. The susceptibilities and the field dependent
magnetic hysteresis loops reveal a strong anisotropy, which could
be modelled in terms of $R$ contributions. The ground state
splitting of the Pr$^{3+}$ ions turns out to be determined by the
crystal field, while in NdMnO$_3$ the splitting is caused by an
effective Nd-Mn exchange field. The ground state splitting of both
compounds is of the order of $\Delta\approx20$~K and could
independently be determined by measurements of the specific heat.
From these studies it follows that the influence of the $R$
sublattice has to be considered as additional driving force of the
puzzling ground state properties of perovskite manganites.

This work was supported by the Bundesministerium
f\"{u}r Bildung und Forschung (BMBF) via VDI/EKM: 13N6917 and by
Sonderforschungsbereich 484 (Augsburg), and by the RFBR
(00-02-16500).




\begin{references}

\bibitem{1} R. von Helmolt, J. Wecker, B. Holzapfel, L. Schultz,
K.~Samwer, Phys. Rev. Lett. {\bf 71}, 2331 (1993); K. Chahara, T.
Ohono, M. Kasai, Y. Kanke, Y. Kozono,   Appl. Phys. Lett. {\bf
62}, 780 (1993).

\bibitem{3} J. M. Coey, M. Viret, S. von Molnar,
  Adv. Phys. {\bf 48}, 167 (1999); E. L. Nagaev,
  Physics Reports {\bf\ 346}, 387 (2001); M. B. Salomon, M. Jaime,
  Rev. Mod. Phys. {\bf 73}, 583 (2001).


\bibitem{par00b} M.~Paraskevopoulos {\em et al.}, J. Magn. Magn. Mater.
{\bf 211}, 118 (2000).

\bibitem{ima98} M.~Imada {\em et al.},
Rev. Mod. Phys. \textbf{70}, 1039 (1998). 

\bibitem{wol55} E.O.Wollan and W.C.Koehler, Phys.Rev. {\bf
100}, 545 (1955).

\bibitem{jon50} G.J.~Jonker and J.H. Van~Santen, Physica \textbf{16},
337 (1950).

\bibitem{wu00} S.Y. Wu {\em et al.},
J. Appl. Phys. \textbf{87}, 5822 (2000).

\bibitem{mun00} A.~Mu\~{n}oz {\em et al.},
J. Phys.: Condens. Mat. \textbf{12}, 1361 (2000).

\bibitem{quez68} S.~Quezel-Ambrunaz, Bull. Soc. Fr. Min\'{e}ral.
Cristallogr. \textbf{B 91}, 339 (1968).

\bibitem{balb} A.M.~Balbashov {\em et al.},
J.~Cryst. Growth. \textbf{167}, 365 (1996).

\bibitem{Cherepanov} V.A. Cherepanov, L.Yu. Barkhatova, A.N.
Petrov, and V.I. Voronin, J. Sol. State Chem. {\bf 118}, 53 (1995)

\bibitem{Kitayama} K.~Kitayama and T.~Kanazaki, J. Sol. State Chem. {\bf 182}, 236
(2001).

\bibitem{Kamehashira} N. Kamehashira  and Y. Miyazaki, Mat. Res. Bull. {\bf 19}, 1201
(1984).

\bibitem{Jirak} Z.~Jir\'ak et al., J. Appl. Phys. {\bf 81}, 5790 (1997).

\bibitem{Ritter} C.~Ritter et al., Phys. Rev. B {\bf 56}, 8902 (1997)).

\bibitem{par00} M.~Paraskevopoulos {\em et al.},
J.~Phys.:~Cond. Mat. \textbf{12}, 3993 (2000).

\bibitem{mukh01} A.A. Mukhin {\em et al.},
J. Magn. Magn. Mater. 226-230, 1139 (2001).

\bibitem{pim00} A. Pimenov {\em et al.},
Phys. Rev. B {\bf 62}, 5685 (2000).

\bibitem{note1} Due to  twins in $a,b$-plane with
interchanged $a$ and $b$ axis the  spin-flop transition in
NdMnO$_3$ as observed  for the two perpendicular directions in the
$a,b$-plane at the same threshold field. Taking into account that
only $H$ is parallel to the $b$-axis could induce a spin-flop type
transition from the magnetic ground state $F_{z}A_{y}$, we were
able to extract pure magnetization along both $a$ and $b$ axis
shown in Fig.~\ref{hysterese} (lower frame).

\bibitem{whit69} R.M. White, J. Appl. Phys. {\bf 40}, 1061 (1969);
 T. Yamaguchi and K. Tsushima, Phys. Rev. B {\bf 8}, 5187 (1973).

\bibitem{note2} We assume here  that the $\bf F$ is eliminated by a standard
minimization procedure and is expressed through dimensionless AFM
moment $\bf A$.

\bibitem{may00} F.~Mayr {\em et al.},
Phys. Rev. B \textbf{62}, 15673 (2000).

\bibitem{mukh03} A.A.~Mukhin {\em et al.},
Phys. Met. Metallogr. {\bf 91},Suppl.1, 194 (2001).

\end{references}
\end{document}